# Einstein's Uniformly Rotating Disk and the Hole Argument

Galina Weinstein

April 22, 2015

Einstein's first mention of the uniformly rotating disk in print was in 1912, in his paper dealing with the static gravitational fields. After the 1912 paper, the rotating disk problem occurred in Einstein's writings only in a 1916 review paper, "The Foundation of the General Theory of Relativity". Einstein did not mention the rotating disk problem in any of his papers on gravitation theory from 1912 until 1916. However, between 1912 and 1914 Einstein invoked the Hole Argument. I discuss the possible connection between the 1912 rotating disk problem and the Hole Argument and the connection between the 1916 rotating disk problem and the Point Coincident Argument. Finally, according to Mach's ideas we see that the possibility of an empty hole is unacceptable. In 1916 Einstein replaced the Hole Argument with the Point Coincidence Argument and later in 1918 with Mach's principle.

In September 1909 Paul Ehrenfest explained in his paper, "Gleichförmige Rotation starrer Körper und Relativitätstheorie" (Uniform Rotation of Rigid Bodies and the Theory of Relativity) that rigid bodies cannot exist in the theory of relativity. In 1909 Max Born discussed the problem of rigid bodies in the theory of relativity and showed that, the concept of the rigid body was incompatible with the special theory of relativity because of the Lorentz contraction. Ehrenfest posed a query which dealt with rotating rigid bodies and the Lorentz contraction in the special theory of relativity:[1]

Consider a relativistic rigid cylinder with radius $R$. It is gradually set into rotation around its axis until it reaches a state of constant rotation with angular velocity $\omega$. As measured by an observer at rest, the radius of the rotating cylinder is $R'$. Then R' has to fulfill the following two contradictory conditions:

1. When the cylinder is moving, an observer at rest measures the circumference of the cylinder to be: $2\pi R' < 2\pi R$ due to the Lorentz contraction of lengths relative to its rest length, since each of its elements moves with an instantaneous velocity $R'\omega$.

2. If one considers each element along the radius of the cylinder, then the instantaneous velocity of each element is directed perpendicular to the radius. Hence the elements of a radius cannot show any contraction relative to their rest length. This means that: $R' = R$.

Ehrenfest had pointed out that a uniformly rotating rigid disk would be a paradoxical object *in special relativity*, since on setting it into motion its circumference would undergo a contraction whereas its radius would remain uncontracted. This is "Ehrenfest's paradox".[2]

On February 28, 1910, Einstein wrote to Vladimir Varićak: "The rotation of the rigid body is the most interesting problem currently provided by the theory of relativity, because the only thing that causes the contradiction is the Lorentz contraction".[3]

Later in 1919, Einstein explained why it was impossible for a rigid disk in a state of rest to gradually set into rotation around its axis. Einstein wrote to the philosopher and mathematician Joseph Petzoldt: [4]

"One must take into account that a rigid circular disk at rest would have to snap when set into rotation, because of the Lorentz shortening of the tangential fibers and the non-shortening of the radial ones. Similarly, a rigid disk in rotation (made by casting) would have to shatter as a result of the inverse changes in length if one attempts to bring it to the state of rest. If you take these facts fully into consideration, your paradox disappears".

In 1912, in Prague, Ehrenfest's paradox, however, motivated Einstein to discuss what came to be known as the uniformly (rigid) rotating disk thought experiment, in the first of two papers on static gravitational fields, "Lichtgeschwindigkeit und Statik des Gravitationsfeldes" (The Speed of Light and the Statics of the Gravitational Fields). Einstein considered a system $K$ with coordinates $x, y, z$ in a state of uniform rotation (disk) in the direction of its $x$-coordinate and referred it to (it is observed from) a non-accelerated system. The origin of $K$ possesses no velocity.[5] Hence, Einstein considered a rigid body *already* in a state of uniform rotation observed from an inertial system.

The system $K$ is equivalent to a system at rest ($K'$) in which there exists a certain kind of mass-free static gravitational field. In 1912 Einstein extended the 1911 equivalence principle to uniformly rotating systems. According to Einstein's 1911 equivalence principle: "if we suppose that the systems $K$ and $K'$ are physically exactly equivalent, i.e., if we assume that we may just as well define the system $K$ as being found in space free from gravitational fields, for we must then consider $K$ as uniformly accelerated".[6] Consequently, "By assuming this, we obtain a principle which, if it is true, has great heuristic meaning. For we obtained by theoretical consideration of the processes which take place relatively to a uniformly accelerating reference system, information as to the course of processes in a homogeneous gravitational field".[7]

Einstein explained that, the special measurements are performed by means of measuring rods. When these are compared with one another in a state of rest at the same location in K, they possess the same length. Hence, the laws of Euclidean geometry must hold for the lengths so measured and for the relations between the coordinates and for other lengths; but "they most probably do not hold in a uniformly rotating system in which, owing to the Lorentz contraction, the ratio of the circumference to the diameter would have to be different from $\pi$ if our definition were applied. The measuring rod as well as the coordinate axes are to be conceived as rigid bodies. This is permitted despite the fact that, according to the [special] theory of relativity, the rigid body cannot really exist. For one can imagine the rigid measuring body being replaced by a great number of non-rigid bodies arranged in a row in such a manner that they do not exert any pressure on each other in that each is supported separately".[8]

Einstein explained the rotating disk problem very succinctly. Even though rigid bodies were not permitted in relativity theory, he still referred to measuring rods as rigid bodies in a coordinate-dependant (static) gravitational theory. "Non-rigid bodies" indicated that Euclidean geometry could not be applied in the rotating disk. We take a great number of small measuring rods (all equal to each other) and place them end-to-end across the diameter 2R and circumference $2\pi R$ of the uniformly rotating disk. From the point of view of a system at rest all the measuring rods on the circumference are subject to the Lorentz contraction (are "non-rigid bodies"). An observer in the system at rest concludes that in the uniformly rotating disk the ratio of the circumference to the diameter is different from $\pi$.

If we adhere to Euclidean geometry, then we arrive at a contradiction because one and the same measuring rod produces what looks like two different lengths due to the Lorentz contraction. From the point of view of a system at rest, the diameter of the uniformly rotating disk would apparently be measured with uncontracted measuring bodies, but the circumference would seem to be longer because the measuring bodies would be contracted. The possibility of Euclidean geometry on a uniformly rotating disk is thus unacceptable.

Neither the measurement of the circumference, nor any other feature of the rotating disk could be said to be in accordance with the Euclidean geometry, for the Euclidean geometry could not be applied in the rotating disk. Thus at one blow the rotating disk ostensibly destroyed the basic merits of the laws of Euclidean geometry, which did not hold anymore for the lengths so measured. Thinking about Ehrenfest's paradox and Born's rigidity and taking into consideration the principle of equivalence, Einstein arrived at the conclusion that a system in gravitational field has non-Euclidean geometry.

Einstein's first mention of the rotating disk in print, as we have said, was in his paper dealing with the static gravitational fields of 1912. However, at about the same time Einstein started to write a manuscript on the special theory of relativity at the request of the Leipzig physicist, Erich Marx, who hoped to have a contribution on relativity theory from Einstein for his *Handbuch der Radiologie*. Although the precise circumstances remain obscure, it appears that Einstein worked on this article off and on in Prague and Zurich from 1912 to 1914, producing a 72-page text.[9] While still in Prague Einstein wrote in the manuscript on the special theory of relativity:[10]

"*Physical meaning of spatial determinations*. The propositions of Euclidean geometry acquire a physical content through our assumption that there exist objects that possess the properties of the basic structures of Euclidean geometry. We assume that appropriately constructed edges of solid bodies not subjected to external influences have the definitional properties that straight lines have (material straight line), and that the part of a material straight line between two marked material points has the properties of a segment. Then the propositions of geometry turn into propositions concerning the arrangements of material straight lines and segments that are possible when these structures are at relative rest".

After the words "and segments that are possible" in the original text, Einstein indicates a note he has appended at the foot of the page: "This term is unclear (rotation)". The reference to rotation is probably an allusion to the problem of the rotating disk discussed in the 1912 paper, "The Speed of Light and the Statics of the Gravitational Fields".[11]

From 1912 onwards, Einstein adopted the metric tensor as the mathematical representation of gravitation. Einstein recognized that the gravitational field should not be described by a variable speed of light as he had attempted to do in his 1912 Prague coordinate dependent theory of static gravitational field. He realized that, the gravitational field is described by a metric, a symmetric tensor-field of metric $g_{\mu\nu}$. The metric tensor field is a mathematical object that characterizes the geometry of space and time. Einstein reasoned that in the general case, the gravitational field is characterized by ten space-time functions of the metric tensor, $g_{\mu\nu}$ are functions of the coordinates ($x_\nu$).

He first found the appropriate mathematical tool used in Minkowski's formalism, the square of the line element, $x^2 + y^2 + z^2 - (ct)^2$, and in tensorial form:

$$(1)\ ds^2 = -\sum_\nu dx_\nu^2$$

which is invariant under the Lorentz group. He then replaced the line element of Minkowski's flat space-time by the more general form:

$$(2)\ ds^2 = \sum_{\mu\nu} g_{\mu\nu} dx_\mu\, dx_\nu$$

In a very small region of space-time the special theory of relativity holds, and $g_{\mu\nu}$ reduce to $g_{44} = c^2$, where c denotes a constant. Einstein took for granted that the same degeneration occurs in the static gravitational field, except that in the latter case, this reduces to a single potential $g_{44} = c^2$, where $g_{44} = c^2$ is a function of spatial coordinates, $x_1, x_2, x_3$. Hence, in this very small region of space-time:[12]

"ds [equation (1)] is the square of the four-dimensional distance between two infinitely close space-time points, measured by means of a rigid body that is not accelerated in" this local system, "and by means of unit measuring rods and clocks at rest relative to it.

From the foregoing, one can already infer that there cannot exist relationships between the space-time coordinates $x_1$, $x_2$, $x_3$, $x_4$ and the results of measurements obtainable by means of measuring rods and clocks that would be as simple as those in the old [special] relativity theory. With regard to time, this has already found to be true in the case of the static gravitational field. The question therefore arises, what is the physical meaning (measurability in principle) of the $x_1$, $x_2$, $x_3$, $x_4$. [...] From this one sees that, for given [differentials] $dx_1$, $dx_2$, $dx_3$, $dx_4$, [... the $ds^2$, equation (2)] that corresponds to these differentials can be determined only if one knows the quantities $g_{\mu\nu}$ that determine the gravitational field. This can also be expressed in the following way: the gravitational field [$g_{\mu\nu}$] influences the measuring bodies and clocks in a determinate manner".

In a 1916 review paper, "Die Grundlage der allgemeinen Relativitätstheorie" (The Foundation of the General Theory of Relativity) (see further below), Einstein applied equation (2) and the above method to the uniformly rotating disk.

After 1912, the rotating-disk argument occurred in Einstein's writings only in the above 1916 review article. Einstein did not mention the rotating disk problem in any of his papers on gravitation theory from 1912 until 1916.[13]

In 1913 Einstein invoked the Hole Argument. According to the Hole Argument, if the field equations are generally covariant, then inside the hole one and the same matter distribution produces what looks like two different metric fields. Perhaps the 1912 disk problem and Ehrenfest's paradox spurred Einstein to invent the Hole Argument: If we adhere to Euclidean geometry, then we arrive at a contradiction because one and the same measuring rod produces what looks like two different lengths due to the Lorentz contraction. From the point of view of a system at rest, the diameter of the uniformly rotating disk would apparently be measured with uncontracted measuring bodies, but the circumference would seem to be longer because the measuring bodies would be contracted. The possibility of Euclidean geometry on a uniformly rotating disk is thus unacceptable.

When and where does the Hole Argument get mentioned for the first time? On November 2, 1913, Einstein was apparently "very happy with the gravitation theory. The fact that the gravitational equations are not generally covariant, which bothered me some time ago, has proved to be unavoidable; it can easily be proved that a theory with generally covariant equations cannot exist if it is required that the field be mathematically completely determined by matter".[14] Einstein "easily proved" that "a theory with generally covariant equations cannot exist" with the Hole Argument.

The Hole Argument was first published in the *Bemarkungen* (added remarks), which forms the addendum to Einstein and Marcel Grossmann's 1913 *Entwurf* paper: *Entwurf einer verallgemeinerten Relativitätstheorie und eine Theorie der Gravitation* (Outline of a Generalized Theory of Relativity and of a Theory of Gravitation).[15]

The Hole Argument reappeared again in Einstein's paper, "Prinzipielles zur verallgemeinerten Relativitätstheorie und Gravitationstheorie" (On the Foundations of the Generalized Theory of Relativity and the Theory of Gravitation), published in January 1914. At this stage Einstein had already become defensive, because his failure to offer generally covariant field equations was a great worry and embarrassment for him.[16] He felt that he had assurance that his theory was founded on deep and precise foundations (the equivalence principle and Mach's ideas), but upon general impression his gravitational field equations failed. He then wrote on a less tentative basis: " 'Very true', thinks the reader, 'but the fact the Messrs. Einstein and Grossmann are not able to give the equations for the gravitational field in generally covariant form is not a sufficient reason for me to agree to a specialization of the reference system'. But there are two weighty arguments that justify this step, one of them of logical, the other one of empirical provenance". The logical argument was the Hole Argument:[17]

Consider a finite portion of space-time where there is no matter. This is the hole. The stress-energy tensor $T_{\mu\nu}$ vanishes inside the hole. $T_{\mu\nu}$, which is outside the hole, therefore, also determines completely everywhere the components of the metric tensor $g_{\mu\nu}$ inside the hole. We now imagine the following substitution: instead of the original coordinate system, $x_\nu$, a new coordinate system $x'_\nu$ is introduced. $x_\nu$ and $x_\nu'$ are different from one another inside the hole $x_\nu \neq x'_\nu$, but outside and on the boundary of the hole they coincide $x_\nu = x'_\nu$. Inside the hole these two coordinate systems lead to two different metric fields $g'_{\mu\nu} \neq g_{\mu\nu}$. Outside the hole: $T'_{\mu\nu} = T_{\mu\nu}$ everywhere, but inside the hole: $T'_{\mu\nu} = T_{\mu\nu} = 0$. Hence, in the case considered, if the above substitutions are allowed, then $T_{\mu\nu}$ determines the components of two different metric tensors inside the hole: $g_{\mu\nu}$ and $g'_{\mu\nu}$. In this way the $g_{\mu\nu}$ cannot be completely determined by the $T_{\mu\nu}$ inside the hole. If the field equations are generally covariant, then inside the hole two different metric fields are produced by the same matter field.

Einstein had persuaded himself that generally covariant field equations are not permissible. Curiously he remained convinced that he had done the right thing. Around March 1914 – Einstein was about to leave Zurich and start his Berlin period. His collaboration with Grossmann was going to end. Before Einstein left he wrote his last joint paper with Grossmann that included just another excuse for not presenting generally covariant field equations, "Kovarianzeigenschaften der Feldgleichungen der auf die verallgemeinerte Relativitätstheorie gegründeten Gravitationstheorie" (Covariance Properties of the Field Equations of the Theory of Gravitation Based on the General Theory of Relativity).[18] The paper was published in May 1914 when Einstein was already in Berlin.

In section §2 of the paper Einstein brought a new "simple consideration" according to which the metric field components that characterize the gravitational field could not be completely determined by generally-covariant equations. He first explained the argument to his close friend Michele Besso on March 10, 1914. From the gravitation equations and the conservation law it follows that there are four coordinate conditions by which one can restrict the coordinate system. For this reason, the gravitation equations hold for every reference system that is adapted to this condition, for *angepaßte Koordinatensysteme* (adapted coordinate systems). The covariance of the field equations was nevertheless far-reaching in these adapted coordinate systems.

On account of the adapted coordinate systems we expect that "there exists acceleration transformations of varied kinds, which transform the equations to themselves (e.g. also rotation), so that the equivalence hypothesis is preserved in its original form, even to an unexpectedly large extent".[19] Einstein was under the impression that the metric field describing space and time for a rotating system was a solution of the field equations of the Einstein-Grossman *Entwurf* theory. Restriction under the four coordinate conditions would apparently solve the problem of physical equivalence of a centrifugal field and a gravitational field, and the *Entwurf* field equations would satisfy the relativity principle and the equivalence principle. Einstein was to discover much later that he made a mistake: the rotation metric was not a solution of the *Entwurf* field equations.

Einstein's "simple consideration" not only justified the restricted covariance of the field equations, but even supplied reasoning for why generally covariant field equations would be unacceptable. In section §2 of the 1914 paper, Einstein reproduced the Hole Argument, and wrote: "Having thus recognized that the useable theory of gravitation requires a necessary specialization of the coordinate system, we also see that the gravitational equations given by us are based upon special coordinate system".[20]

With the Hole Argument the theory acquires a necessary limitation, but with the adapted coordinate systems "*the gravitational equations [...] are covariant with respect to all admissible transformations of the coordinate systems, i.e., with respect to all transformations between coordinate systems which satisfy the [four] conditions*". And the proof for this claim was that, since the conditions by which one restricted the coordinate systems are direct consequence of the gravitational equations, therefore the covariance of the equations is far-reaching.[21] The four conditions were to hold in all those systems, which were adapted coordinate systems and in which the *Entwurf* field equations were valid. Einstein thus posed a coordinate condition on the field equations and the coordinate systems satisfying this condition were the adapted coordinate systems for the gravitational field. Coordinate transformations between two such adapted coordinate systems were arbitrary non-linear transformations.[22] However, the covariance of Einstein's equations was only "far-reaching".

Einstein stubbornly adhered to his adapted coordinate systems, and could not hear any criticism. In fact, he had a propensity for clinging to ideas (e.g. the adapted coordinate systems and later the static cosmological model), even after they have been mathematically discredited.

By October 1914, Einstein completed in Berlin a summarizing long review article on his *Entwurf* theory, published in November that same year. In this paper Einstein presented the most comprehensive, organized and extended version of his *Entwurf* theory.

In section §1, "Introductory Considerations", Einstein considered a version of Newton's bucket experiment. Consider two systems: one K, which is in uniform translation motion, and is a Galilei-Newtonian coordinate system, and the other K', which is in uniform rotation relative to K. Centrifugal forces act on the masses at rest relative to K', while they do not act upon the masses which are at rest relative to K. Newton already saw in this a proof that the rotation of K' had to be regarded as "absolute", and thus one could not consider K' as "at rest" like K. This argument, however, wrote Einstein – as shown particularly by Ernst Mach – is not valid. The existence of these centrifugal forces do not necessarily require the motion of K'. We could just as well derive them from the averaged rotational movement of distant masses in the environment with respect to K', and thereby treating K' as "at rest".[23]

Einstein was under the impression that the above argument spoke in favor of his new theory of general relativity: the centrifugal force, which acts under given conditions upon a body, is determined by precisely the same natural constant as the effect of the gravitational field, such that we have no means to distinguish a "centrifugal field" from a gravitational field: the physical equivalence of a centrifugal field and a gravitational field. He thus interpreted the rotating system K' as *at rest* and the centrifugal field as a gravitational field, and introduced *the inertio-gravitational field*.

Einstein had again become rigid and defensive and explained: it appears the necessary demand is that the differential laws of the metric tensor components must be generally-covariant, but "we want to show that we have to restrict this demand if we want to fully satisfy the theorem of causality. We shall prove, namely, that it can be impossible that the laws that determine the course of events in a gravitational field are generally covariant". And the "Proof of Necessary of a Restriction of the Choice of Coordinates" was presented in section §12: an improved and elaborated version of a Hole Argument.[24] Consider a finite part of space-time (hole) where material processes do not occur. Physical events are completely determined if the metric field components $g_{\mu\nu}$ are functions of $x_\nu$ with respect to a coordinate system K. The totality of the $g_{\mu\nu}$ is symbolically denoted by G(x). Consider the gravitational field with respect to a new coordinate system K' and the totality of the $g'_{\mu\nu}$ ($x_\nu'$) is symbolically denoted by G'(x'). "G'(x') and G(x) describe the same gravitational field" outside the hole, but smoothly differ one from another inside the hole. We now make the following substitution: we replace the coordinate $x'_\nu$ by the coordinate $x_\nu$ in the functions $g'_{\mu\nu}$ for the region inside the hole. We thus obtain: G'(x). Hence, G'(x) also describes a gravitational field with respect to a new coordinate system K inside the hole. If we assume now the differential equations of the gravitational field are generally covariant, then they are satisfied for G'(x') (with respect to K') if they are satisfied by G(x) with respect to K. They are then also satisfied with respect to K by G'(x). With respect to K there then exist *two* different solutions G(x) and G'(x), which are different from one another, nevertheless at the boundary of the region both solutions coincide, *i.e., what is happening cannot be determined uniquely by generally-covariant differential equations for the gravitational field*".

Einstein demonstrated that the Hole Argument was compatible with the restriction of the coordinate systems to adapted coordinate systems. He considered a finite portion of space-time (a hole) and the coordinate system K, and imagined a series of infinitely related coordinate systems K', K". Einstein demonstrates that among these systems there are "systems adapted to the gravitational field". The four coordinate conditions (that are direct consequence of the gravitational equations) hold for these adapted systems in which two metric field components $g'_{\mu\nu}$ and $g_{\mu\nu}$ are not functions of the same $x_\nu$. This is a sufficient condition that the coordinate system is adapted to the gravitational field and that the covariance of the field equations is far-reaching in these adapted coordinate systems.[25]

The adapted coordinate systems and Einstein's gravitational tensor turned out to be so implausible that people were inclined to reject his gravitational tensor. A year later in October 1915, Einstein wrote to Paul Hertz: "But I have shown in my paper that a usual gravitation law cannot be generally covariant. Do you not agree with this consideration?"[26] Einstein apparently had not yet renounced the *Entwurf* field equations when writing this, but he probably had serious doubts, because he was extremely nervous and cynical. Later during October 1915 Einstein silently dropped the Hole Argument and did not mention it in his November 1915 general relativity papers.

Einstein came to believe the metric field describing a rotating system was a solution of his field equations, and he was led to think that the *Entwurf* field equations hold in rotating frames. Einstein discovered only around October 1915 that the metric field describing a rotating system was not a solution of his *Entwurf* field equations and wrote in the oft-quoted letter to Arnold Sommerfeld:[27] "I realized, namely, that my existing field equations of gravitation were entirely untenable! […] I proved that the gravitational field on a uniformly rotating system does not satisfy the field equations".

In December 1915 Ehrenfest posed a new query. He asked Einstein about the Hole Argument from section §12 of his 1914 review article. Einstein silently dropped the Hole Argument during October 1915 and did not mention it in his November 1915 papers, and had to explain to Ehrenfest just what was wrong with the Hole Argument against general covariance, in which he believed so strongly for over two years.

The reason why Einstein dropped the Hole Argument, while he adopted a new argument is that in section §12 of his paper of "last year, everything was correct (in the first 3 paragraphs) up to the italicized part at the end of the third paragraph". Further, in the Hole Argument G(x) and G'(x) represented two different gravitational fields with respect to the same reference system. *The correction should be that the reference system has no meaning and that the realization of two gravitational fields in the same region of the continuum is impossible*". It therefore be thought that in general relativity "The following consideration should replace §12". Einstein assigned great significance to the new "consideration". The consideration, the Point Coincidence Argument, was later presented in section §3 of the 1916 review paper, "The Foundation of the General Theory of Relativity":[28]

"The hole argument is replaced by the following consideration. Nothing is physically *real* but the totality of space-time point coincidences. If, for example, all physical events were to be built up from the motions of material points alone, then the meetings of these points, i.e., the points of intersection of the world lines, would be the only real things, i.e., observable in principle. These points of intersection naturally are preserved during all [coordinate] transformations (and no new ones occur) if only certain uniqueness conditions are observed. It is therefore most natural to demand of the laws that they determine no *more* than the totality of space-time coincidences. From what has been said, this is already attained through the use of generally covariant equations".

Einstein thus avoided the Hole Argument quite naturally by virtue of the exercise of the deep insight that there is no meaning to space-time measurements: all points are indistinguishable from one another. In the 1914 Hole Argument, considering generally covariant field equations, then with respect to K there exist two different solutions G(x) and G'(x), which are different from one another, but at the boundary of the hole both solutions coincide. However, according to the Point Coincidence Argument, G(x) and G'(x) should represent the same gravitational field, because all events consisted only of the motion of material points, and we are dealing only with the coincidences of the space-time points. Suppose we take the first solution – gravitational field – G(x) and think of it as material points moving in space-time. Then we are dealing only with the meeting of the two material points. Let us designate this meeting point by $(x_1, x_2, x_3, x_4)$. We shall take the second solution G'(x) and think of it as well as material points moving in space-time. This solution, according to Einstein's Point Coincidence Argument, also reduces to the meeting of two material points at $(x'_1, x'_2, x'_3, x'_4)$. The two points are indistinguishable, because there is no reason to preferring the first to the latter.

In section §3 of the 1916 review paper, before presenting the Point Coincidence Argument, Einstein deals with the uniformly rotating disk problem. In 1915 Einstein realized that coordinates of space and time have no direct physical meaning in general relativity because "time and space are deprived of the last trace of objective reality".[29] Fairly soon after this, Einstein reconsidered the 1912 rotating disk thought experiment in section §3 of the 1916 review paper, "The Foundation of the General Theory of Relativity".

Einstein explained that *in classical mechanics and in special relativity*, space-time measurements of four-coordinates are done *with rods and standard clocks*. With these we define lengths and times in all inertial reference frames. The notions of coordinates and measurements in classical mechanics and in special relativity *presuppose the validity of Euclidean geometry*. Also according to the special theory of relativity, the laws of geometry are directly interpreted as laws relating to the possible relative positions (at rest) of solid bodies, and, the *laws of kinematics* are to be interpreted as laws which describe the relations of measuring bodies and clocks. "To two selected material points of a stationary (rigid) body there always corresponds a distance of quite definite length, which is independent of the location and orientation of the body, as well as of the time; to two selected positions of the hands of a clock at rest relatively to the (legitimate) reference system, there always corresponds an interval of time of a definite length, which is independent of place and time".

Thus the general theory of relativity now emerged as a new theory: "It will soon be seen that the general theory of relativity cannot adhere to this simple physical interpretation of space and time".[30] Moreover, in the general theory of relativity the method of laying coordinates in the space-time continuum (in a definite manner) breaks down, and one cannot adapt coordinate systems to the four-dimensional space. This seemed the greatest breakthrough made by Einstein about space-time measurements.

In view of what was said above, Einstein considered two systems, a Galilean system K, and the other K', which is in uniform rotation relative to K. He then demonstrated by the rotating disk thought experiment that we are unable to define properly coordinates in K', and Euclidean geometry breaks down for K'. He concluded that we are also unable to properly define time by clocks at rest in K' either. The coordinates of space and time, therefore, have no direct physical meaning with respect to K'.

The origin of both systems, as well as their axes of z, permanently coincide one with another. The circle around the origin in the x, y plane of K is regarded at the same time as a circle in the x', y' plane of K'. In section §3 of Einstein's 1916 review paper the uniformly rigid rotating body became a *circle*: "We now imagine that the circumference and diameter of this circle are measured with a unit measure (infinitely small relative to the radius), and we form the quotient of the two results". If the experiment is performed with a measuring rod at rest relative to K, the quotient will be $\pi$. With a measuring rod at rest relative to K', the quotient will be greater than $\pi$. This can be seen, if the whole process of measurement is viewed from the system K, taking into consideration that the periphery undergoes a Lorentz contraction, while the measuring rod applied to the radius does not.[31]

12An observer in a Galilean system K is an inertial observer and he, therefore, uses special relativity to measure the circumference and diameter of the circle. In 1909 Ehrenfest suggested that in a uniformly rotating disk the Lorentz contraction causes the circumference to *be shorter* than a circumference in a disk at rest. Naturally, if we consider Einstein's explanation we find, of course, quite the opposite. The measurement is performed from the Galilean system K. The circumference of the uniformly rotating disk, therefore, appears *longer* than $2\pi R$ as measured by an observer in K. A remark of Einstein to Petzoldt explains this matter:[32] "Now, you think that the rigidly rotating circular line would have to have a circular circumference that is less than $2r\pi$, due to the Lorentz contraction. It is here that the root of the error lies, that you instinctively set the radius $r$ of the rotating circular line equal to the radius $r_0$ that the circular line has in the stationary case. This is incorrect, however; because, owing to the Lorentz contraction, $2\pi r = 2\pi r_o \sqrt{1 - \frac{v^2}{c^2}}$ ".

Subsequently Einstein treated time measurements. He did not explicitly discuss this matter in his 1912 paper, "The Speed of Light and the Statics of the Gravitational Fields".

In his 1907 paper, "Über das Relativitätsprinzip und die aus demselben gezogenen Folgerungen" (On the Relativity Principle and the Conclusions Drawn from It), Einstein considered a reference system S' that is uniformly accelerated relative to a non-accelerated system S in the direction of its x-axis. The clocks of S' are set at time t', and he asked what is the rate of the clocks in the next time element t?[33] According to the 1907 equivalence principle, time measurements in a uniformly accelerated reference system are also valid for a coordinate system in which a homogeneous gravitational field is acting. Consequently, for an observer located somewhere in space, the clock in a gravitational potential runs faster than an identical clock located at the coordinate origin.[34] In 1911, Einstein reconsidered the measurement of time in uniform accelerated systems, "If we measure time in $S_1$ with a clock U, *we must measure the time in $S_2$ with a clock that goes [...] slower than the clock U if you compare it with the clock U in the same place*".[35]

In 1916 Einstein extended the 1907 and 1911 time measurements to time measurements in uniformly rotating systems. Einstein imagined two clocks of identical constitution placed, one at the origin of coordinates and the other at the periphery of the circle. Both clocks are observed from the Galilean system K. "According to a known result from special relativity – judged from K – the clock at the periphery of the circle goes more slowly than the other clock at the origin, because the clock at the former [the circumference] is in motion and the latter [at the origin] is at rest". An observer who is located at the origin, and who is capable of observing the clock at the circumference by means of light, would be able to see the periphery clock lagging behind the clock beside him. He will interpret this observation as showing that the clock at the periphery "really" goes more slowly than the clock at the origin.

He will thus define time in such a way that the rate of the clock depends upon its location.[36]

Clocks on the uniformly rotating disk cannot be synchronized according to Einstein's synchronization method from his 1905 special relativity paper: in special relativity we synchronize two clocks, B with clock A, by sending a light signal from A to B and then back from B to A. Therefore, measuring rods and clocks cannot be the fundamental entities in the general theory of relativity.

Einstein arrived at two important results: when we measure the circumference of the circle of K' from the system K, "the measuring-rod applied to the periphery undergoes a Lorentzian contraction, while the one applied along the radius does not"; and when we require measurement of time events in K', then judged from K, "the clock at the periphery of the circle goes more slowly than the other clock at the origin".

As before, Einstein explained the matter to Petzoldt: "You are making the same analogous error for clocks as for measuring rods. The *rotating observer notices very well that, of both his identical clocks, the one positioned at the circumference runs more slowly than the one positioned at the center*".[37]

Einstein concluded *that we cannot define properly coordinates in K'. The lengths measurements have no direct meaning for K'. We are unable to properly define time by clocks at rest in K' either. The coordinates of space and time, therefore, have no direct physical meaning with respect to K'.* [38]

Einstein returned to the rotating disk thought experiment towards the end of his 1916 paper. According to the 1912 equivalence principle, the system K', which is in uniform rotation relative to the Galilean system K, is equivalent to a system at rest in which there exists a certain kind of mass-free static gravitational field.

Einstein used the first order approximate solution to his vacuum field equations in order to demonstrate that a gravitational field changes the dimensions of measuring rods and the clock period. He considered (2) and small unit-measuring rods laid on the x-axis. Therefore, $ds^2 = -1$ and the differentials $dx_2 = dx_3 = dx_4 = 0$. The first order approximate solution to the vacuum field equations gives:[39]

(3) $g_{11} = -\left(1 + \dfrac{2GM}{c^2 r}\right).$

G – gravitational constant, M – mass of central body, and:

(4) $-1 = g_{11} dx_1^2 = -\left(1 + \dfrac{2GM}{c^2 r}\right) dx_1^2.$

Equations (3) and (4) both yield:

$$(5)\ dx = 1 - \frac{GM}{c^2 r}.$$

Einstein concluded: "The unit measuring rod therefore appears a little shortened with respect to the coordinate system by the presence of the gravitational field, if it is laid in the radial direction".

As to the length of a measuring rod in the tangential direction: we set $ds^2 = -1$, and the differentials: $dx_1 = dx_3 = dx_4 = 0$; $x_2 = r$, $x_1 = x_3 = 0$. Therefore:

$$(6)\ -1 = g_{22} dx_2^2 = -dx_2^2.$$

With the tangential position, therefore, the gravitational field of a point mass has no influence on the length of a rod.

Thus we see that general relativity may be regarded as a theory in which: "Space and time cannot be defined in such a way that spatial coordinate differences be directly measured by the unit measuring rod, and time by a standard clock". The method of laying coordinates in the space-time continuum in a definite manner breaks down, and one cannot adapt coordinate systems to the four-dimensional space.

This conclusion brought Einstein to a formulation of a principle of general covariance: If we cannot be dependent on space and time measurements, then we must regard all imaginable systems of coordinates, on principle, as equally suitable for the description of nature: "*The general laws of nature have to be expressed by equations which are valid for all coordinate systems, i.e., are covariant with respect to any substitutions (generally covariant)*". The argument that supported this principle of general covariance, "which takes away from space and time the last remnant of physical objectivity, can be seen from the following considerations. All our space-time verifications invariably amount to a determination of space-time coincidences". Einstein then presented the Point Coincidence Argument *that replaced the Hole Argument*.[40]

The point being made in the 1916 Point-Coincidence Argument is, briefly, that unlike general relativity, in special relativity coordinates of space and time have direct physical meaning. A simpler way of stating this is to say: We associate to the world four space-time variables $x_1$, $x_2$, $x_3$, $x_4$. For every point-event there is a corresponding system of values of the variables $x_1…x_4$. We associate the above variables, coordinates, to space-time coincidences of point-events. "Two coincident point-events correspond to the same system of values of the variables $x_1…x_4$, i.e., the coincidence is characterized by the identity of the coordinates. If, instead of the variables $x_1…x_4$, we introduce functions of them, $x'_1$, $x'_2$, $x'_3$, $x'_4$, as a new coordinate system, so that the system of values corresponds to one another unambiguously, then the equality of all four coordinates in the new system will also serve as an expression for the space-time coincidence of the two point-events. Since all our physical experience can be ultimately reduced to such point coincidences, there is no immediate reason for preferring certain systems of coordinates to others, i.e., we arrive at the requirement of general covariance".[41]

After presenting the Point Coincidence Argument Einstein very likely intended to include another section under the title "§4 Die fundamentale Mass-Eigenschaft" (§4 The Fundamental property of Mass), but he regretted. In the manuscript of the 1916 paper he wrote the following title on page 7, "§4 The Fundamental property of Mass", and crossed out this sentence.[42]

It is clear that we cannot infer very much from this title about Einstein's intensions. But it is appropriate to ask whether in section §4 Einstein could have thought of formulating an additional argument or principle to replace the Hole Argument. Eventually, the Hole argument is replaced by two principles: the Point Coincidence Argument and Mach's principle. In 1916 Einstein did not yet distinguish between the Point Coincidence Argument and Mach's ideas (inertia should be derived from an interaction of bodies). In 1918 Einstein saw the need to define the principles on which general relativity was based. In his paper, "Prinzipielles zur allgemeinen Relativitätstheorie" (Principles of the General Theory of Relativity), he wrote that his theory rests on three principles, which are not independent of each other. He formulated the principle of relativity in terms of the Point Coincidence Argument and added Mach's principle. The three principles are: [43]

"a) *Relativity Principle*: The laws of nature are merely statements about space-time coincidences; they therefore find their only natural expressions in generally covariant equations.

"b) Equivalence Principle: Inertia and weight are identical in nature. It follows necessarily from this and from the result of the special theory of relativity that the symmetric 'fundamental tensor' [$g_{\mu\nu}$] determines the metrical properties of space, the inertial behavior of bodies in it, as well as gravitational effects. We shall denote the state of space described by the fundamental tensor as the 'G-field'."

"c) *Mach's Principle*[1)]: The G field is *completely* determined by the masses of the bodies. Since mass and energy are identical in accordance with the results of the special theory of relativity and the energy is described formally by means of the symmetric energy tensor ($T_{\mu\nu}$), this means that the G-field is conditioned and determined by the energy tensor of the matter.

1) Hitherto I have not distinguished between principles (a) and (c), and this was confusing. I have chosen the name 'Mach's principle' because this principle has the significance of a generalization of Mach's requirement that inertia should be derived from an interaction of bodies".

John Stachel has written lately that the Hole Argument purports to show that, if the field equations are generally covariant, then Mach's principle cannot be satisfied: Even if the field and all its sources outside of and on the boundary of the hole are specified, such equations cannot determine a unique field in the hole. Hence, the gravitational field equations cannot be generally covariant.[44]

Between 1913 and 1914 Einstein appeared to have also been influenced by Mach when giving several formulations to the Hole Argument. This, however, can be entirely reasonable from the standpoint of the Hole Argument, for the whole tenor of the "hole" is that of a region of space-time empty of matter. Basically Einstein considered "a finite portion of the continuum" (a hole), "in which a material process does not occur" and the inertia of bodies seems not to be caused. For instance, he wrote in the added remarks to the *Entwurf* paper: "Let there be in our four dimensional manifold a portion L, in which a 'material process' is not occurring, and therefore the [components of the stress-energy tensor] $\Theta_{\mu\nu}$ vanish.[45]

The 1916 title, "§4 Die fundamentale Mass-Eigenschaft" (§4The Fundamental property of Mass), has an interesting bearing upon the development of Einstein's Hole Argument. It appears that Einstein himself was forced to restrict the choice of the reference system because in 1914, he believed his theory satisfied Mach's ideas of inertia having its origin in the interaction between mass and all of the other masses, i.e., explained rotation and Newton's bucket experiment. This, according to Einstein, eliminated the epistemological weakness of Newtonian mechanics – the absolute motion. But Einstein's hole, "a finite portion of the continuum, in which a material process does not occur" or "a portion L, in which a 'material process' is not occurring" unraveled the underlying problematic nature of the *Entwurf* theory. As already mentioned, the *Entwurf* theory could not explain rotation and Newton's bucket experiment.

A hole "in which a material process does not occur" may become singular, and according to Mach's ideas we see that the possibility of such a singularity, an empty hole, is unacceptable. Einstein thus replaced the Hole Argument with the Point Coincidence Argument and later in 1918 with Mach's principle.

---